\documentclass[showpacs,amsmath,amssymb, nobibnotes, aps, prl,showkeys]{revtex4-2}
\usepackage{graphicx}
\usepackage{dcolumn}
\usepackage{bm}
\usepackage{docs}
\usepackage{boldline,multirow}
\usepackage[dvipsnames]{xcolor}
 \usepackage{lineno}
\expandafter\ifx\csname package@font\endcsname\relax\else
 \expandafter\expandafter
 \expandafter\usepackage
 \expandafter\expandafter
 \expandafter{\csname package@font\endcsname}%
\fi

\begin{document}
\title[High energy leptonic originated neutrinos from astrophysical objects]{High-energy leptonic originated neutrinos from astrophysical objects}

\author{Arunava Bhadra}
\affiliation{High Energy \& Cosmic Ray Research Centre, University of North Bengal, Siliguri, 734013, India } 
\email{aru\_bhadra@yahoo.com}

\author {Prabir Banik} 
\affiliation{ High Energy \& Cosmic Ray Research Center, University of North Bengal, Siliguri WB 734013, India. }
\email{pbanik74@yahoo.com}


\begin{abstract}
High-energy neutrinos are traditionally regarded as unambiguous signatures of hadronic cosmic rays in astrophysical environments. Here we show that TeV neutrinos can instead be produced by energetic electrons through purely electromagnetic processes in a variety of potential cosmic-ray accelerators. The resulting fluxes are comparable to those expected from hadronic interactions, suggesting that electrons may contribute a significant fraction of the neutrinos detected by the IceCube Observatory. These findings challenge the conventional interpretation of neutrino origins and underscore the need for joint gamma-ray and neutrino observations over a broad energy range to discriminate between hadronic and leptonic production mechanisms.

\end{abstract}


\maketitle

\section{Introduction}
The discovery of extraterrestrial high-energy cosmic neutrinos (with energies of about $20$ TeV and above) by the IceCube Observatory in 2012 \cite{ab1,ab1a,ab1b,ab1c,ab1d} marked the advent of high-energy neutrino astronomy. Such observations are widely regarded as a key to identifying the acceleration sites of hadronic cosmic rays, thereby addressing one of the long-standing questions in high-energy astrophysics \cite{ab2,ab2a,ab2b,ab2c,ab2d,ab2e}. Energetic hadrons interacting with ambient matter or radiation fields produce pions, which subsequently decay into neutrinos and gamma rays. As both particles are electrically neutral, they can travel unhindered from their sources to Earth, serving as unique messengers for locating and characterizing cosmic-ray accelerators. However, very-high-energy gamma rays can also originate from relativistic electrons through inverse Compton scattering, meaning that their detection alone cannot definitively establish hadronic acceleration at the source. In contrast, high-energy neutrinos are generally assumed to arise predominantly from hadronic interactions \cite{ab2,ab2a,ab2b,ab2c,ab2d,ab2e}.

In this work, we show that high-energy astrophysical neutrinos can also be produced by very energetic electrons through electromagnetic interactions in a variety of potential cosmic-ray sources, yielding fluxes comparable to those expected from hadronic processes at very high energies (as quantified in a later section). This finding implies that the detection of very-high-energy neutrinos alone may not conclusively identify hadronic cosmic-ray sources. We argue that only a combined analysis of high-energy gamma rays and neutrinos—revealing consistent flux levels and spectral features across a broad energy range—can unambiguously determine whether their origin is hadronic or leptonic.

The idea of neutrino production from leptonic processes was probably first proposed by Kusenko and Postma \cite{ab3}, who suggested that ultrahigh-energy photons, possibly originating from certain classes of topological defects, could generate ultrahigh-energy electrons through interactions with the cosmic microwave background radiation (CMBR) at high redshifts via double pair production \cite{ab3}. These electrons, in turn, can produce neutrinos by scattering off the CMBR through the muon pair production process. At relatively low energies, inverse Compton (IC) scattering ($e + \gamma \rightarrow e + \gamma$) converts high-energy electrons into high-energy photons. Beyond the electron–positron pair production threshold, higher-order processes such as triplet pair production (TPP; $e + \gamma \rightarrow e + e^{+} + e^{-}$) dominate. At even higher energies—above the muon or pion pair production thresholds—electron interactions with ambient photons can produce a muon–electron pair (MPP; $e + \gamma \rightarrow e + \mu^{+} + \mu^{-}$) or a charged pion pair (PPP; $e + \gamma \rightarrow e + \pi^{+} + \pi^{-}$) through third-order electromagnetic processes. The secondary muons and pions subsequently decay into neutrinos.

Using the relevant Feynman's diagrams, the differential cross-section of the MPP process in the leading order as a function of the outgoing muon energy $E_{\mu}$, in head on collisions, can be expressed in terms of a convolution expression \cite{ab4}:
\begin{eqnarray}
\label{eq1}
 \frac{\mbox{d}\sigma_{MPP}}{\mbox{d}y}\, &\simeq& \frac{\alpha^{3}}{m_{\mu}^2y} \left[1+\left(1-\frac{4m_{\mu}^2y^2}{s}\right)^2\right]
 \ln{\left(\frac{s}{m^{2}_{e}}  \right)}\nonumber \\
&\times&(1-v^2)\left[\left(1+\frac{1}{y^2}\right) \ln{\left(\frac{1+v}{1-v}\right)} 
-v-\frac{1}{y^4}\left(\frac{v}{1-v}+\tanh^{-1}(v)\right)\right],
\end{eqnarray}
where $y=E_{\mu}/m_{\mu}$ with $y_{min}=1$ and $y_{max}=\sqrt{s}/2m_{\mu}$; $v$ is the velocity of the outgoing muon, which is related to $y$ by $v=\sqrt{1-\frac{1}{y^{2}}}$. Here $\alpha$ ($\simeq 1/137$) is the electromagnetic fine-structure constant, and $m_{\mu}$ and $m_{e}$ are the rest mass of the muon and electron respectively. The total cross section $\sigma_{MPP}$ is obtained by performing the $y$ integration. Replacing $\mu^{\pm}$ by $e^{\pm}$ and $\pi^{\pm}$ in the convolution expression \ref{eq1}, one obtains the cross-sections of TPP and PPP respectively.   

In the CM frame the inelasticity for MPP process is defined as

\begin{eqnarray}
\label{inelasticity}
 \eta_{MPP}(s)=\frac{1}{\sigma_{MPP}(s)}\int dE_{\mu}  \left(\frac{E_{\mu}}{\sqrt{s}}\right)  \frac{d\sigma_{MPP}}{dE_{\mu}}.
\end{eqnarray}

The inelasticity for TPP in the large $s$ limit is given by $\eta_{TPP} \sim 3.44 \left(s/m_{e}^{2}\right)^{-0.5}$ (in the CM frame) \cite{ab4}. 

Kusenko and Postma \cite{ab3} assumed that the interaction length for MPP exceeds the energy attenuation length due to triplet pair production when $s > 5m_{\mu}^2$, making MPP an efficient mechanism for generating high-energy cosmic neutrinos. Mathematically, MPP becomes effective when the ratio $R > 1$, where 

\begin{eqnarray}
\label{R}
R \equiv \frac{\sigma_{MPP}(s)}{\eta_{TPP}(s)\,\sigma_{TPP}(s)},
\end{eqnarray}

and $\eta_{TPP}(s)$ is the inelasticity for TPP in the laboratory frame. They estimated $R \sim 100$ in their scenario. However, Athar et al. \cite{ab4} showed that the correct interaction cross-section for MPP in the energy range $5m_{\mu}^2 \lesssim s \lesssim 20m_{\mu}^2$ is about three orders of magnitude smaller than that adopted by Kusenko and Postma \cite{ab3}, leading to $R \ll 1$ in their case. Consequently, the interaction length for MPP is much larger than the energy attenuation length due to TPP, rendering MPP inefficient for high-energy neutrino production at high redshifts.

In the following, we argue that under certain astrophysical conditions or within specific source environments, the ratio $R$ can indeed exceed unity. Moreover, even when $R < 1$, if the threshold condition for MPP is satisfied, neutrinos will still be produced—albeit with fluxes substantially lower than in the $R > 1$ regime—but still relevant for interpreting current observations.
 
\section*{Results}

Electrons are accelerated to very high energies in a variety of astrophysical environments. The maximum energy of accelerated electrons in supernova remnants (SNRs), inferred from their non-thermal X-ray and $\gamma$-ray emissions, is typically around 100~TeV \cite{ab5a,ab5b}. The detection of bright, day-long $\gamma$-ray flares above 100~MeV in the Crab Nebula by the space telescopes \textit{AGILE} \cite{ab5} and \textit{Fermi} \cite{ab6}—attributed to synchrotron radiation from relativistic electron–positron pairs—indicates the presence of PeV electrons in the nebula. A multi-wavelength study of a sample of sources detected by the Large High Altitude Air Shower Observatory (LHAASO), which has revealed several PeV emitters, further suggests that electrons can be accelerated to PeV energies in pulsar wind nebulae \cite{Joshi23}. Microquasars, believed to be powered by rapidly spinning black holes with masses up to a few tens of solar masses, may also serve as potential sites for PeV electron acceleration \cite{LHAASO24}. Moreover, in extragalactic cosmic-ray sources such as active galactic nuclei (AGNs) and gamma-ray bursts (GRBs)—both widely recognized as capable of accelerating particles to energies far beyond those attainable in typical Galactic accelerators—the maximum electron energies inferred from their broadband spectral energy distributions (SEDs) likewise reach the PeV range \cite{Tavecchio11, Aharonian08, Sudoh20}. In shearing / large-scale jets of AGN, electrons up to several
PeV may be achievable under favorable conditions \cite{Rieger19}.

Balancing the acceleration rate with synchrotron and inverse-Compton losses, the maximum energy of an electron accelerated under stochastic (second-order Fermi) acceleration is given by (e.g. \citep{Khangulyan08, Reynolds98}):
\begin{equation}
E_{max} \simeq 630 \left(\frac{B}{1 \; mG} \right)^{-1/2} \left(\frac{\eta}{10} \right)^{-1/2}  \; TeV,
\end{equation}
where B is the magnetic-field strength in Gauss. and the parameter $\eta$ quantifies the acceleration efficiency. This relation implies that in an efficient accelerators ($\eta \rightarrow 5$), electrons can reach PeV energies or beyond when the magnetic field is in the milli-Gauss range or weaker. In AGN environments, magnetic field strengths vary significantly depending on the spatial region. Hotspots — where jets terminate and form strong shocks—typically exhibit fields of a few micro-Gauss, based on synchrotron minimum-energy estimates (e.g., \citep{Hardcastle1998, Worrall2006}). In contrast, the inner relativistic jet on parsec scales often hosts milli-Gauss-level fields, inferred from VLBI core-shift measurements and synchrotron self-absorption modeling (e.g., \citep{O’Sullivan2009, Zamaninasab2014}). For comparison, typical magnetic-field strengths in supernova remnants range from tens to hundreds of micro-Gauss, and may reach milli-Gauss levels in young or strongly shocked regions (e.g., \citep{Vink2012, Uchiyama2007}). Under such conditions, electrons can in principle attain energies approaching the PeV scale, consistent with observations of Galactic PeVatrons.

The threshold energy ($\epsilon_{e,th}$) for the muon pair production (MPP) reaction, expressed in natural units, is $\epsilon_{e,th} = \sqrt{s_{th}} > 2m_{\mu} + m_{e}$, which translates to the condition $\epsilon_{e} \epsilon_{\gamma} > 0.02 f_{g} \; GeV^{2} $ where $f_{g} = (1 - \cos \theta_{e\gamma})^{-1}$, $\epsilon_{e}$ and $\epsilon_{\gamma}$ denote the electron and photon energies, respectively, and $\theta_{e\gamma}$ is the angle between the electron and photon directions. For the pion pair production (PPP) reaction, the right-hand side of this inequality is replaced by $2m_{\pi}$, making the threshold energy slightly higher than that for MPP. If $\epsilon_{e}$ is of the order of a PeV, $\epsilon_{\gamma}$ must be a few tens of eV to satisfy the threshold condition. The thermal plasma (shock-heated) in AGN hotspots typically reaches temperatures of $10^7$ – $10^8$ K \citep{Croston2007, Wilson2006}, corresponding to characteristic energies of roughly 1–10 keV. The typical temperatures of accretion disks in AGNs \cite{ab7,ab7a}, young supernova remnants \cite{ab8, ab8a}, and young neutron stars \cite{ab9, ab10} are approximately $0.1$~keV (corresponding to $10^{6}$~K), while in GRBs the temperatures are even higher \cite{ab11, ab11a, ab11b}. Thus, in all these environments, the threshold condition for the MPP can be readily satisfied, provided that electrons are accelerated to PeV energy range . 

A relevant question at this stage is whether the MPP interaction is an efficient mechanism for generating high-energy astrophysical neutrinos. Both Kusenko and Postma \cite{ab3} and Athar et al. \cite{ab4} assessed the effectiveness of the MPP process for neutrino production through the parameter $R$. In Figure~1, we plot the variation of $R$ as a function of $s$. The ratio $R$ exceeds unity around $s \approx 650 m_{\mu}^{2}$, suggesting that MPP could indeed be a viable mechanism for high-energy neutrino production in the aforementioned environments.

\begin{figure}[h]
  \begin{center}
\includegraphics[scale=0.45]{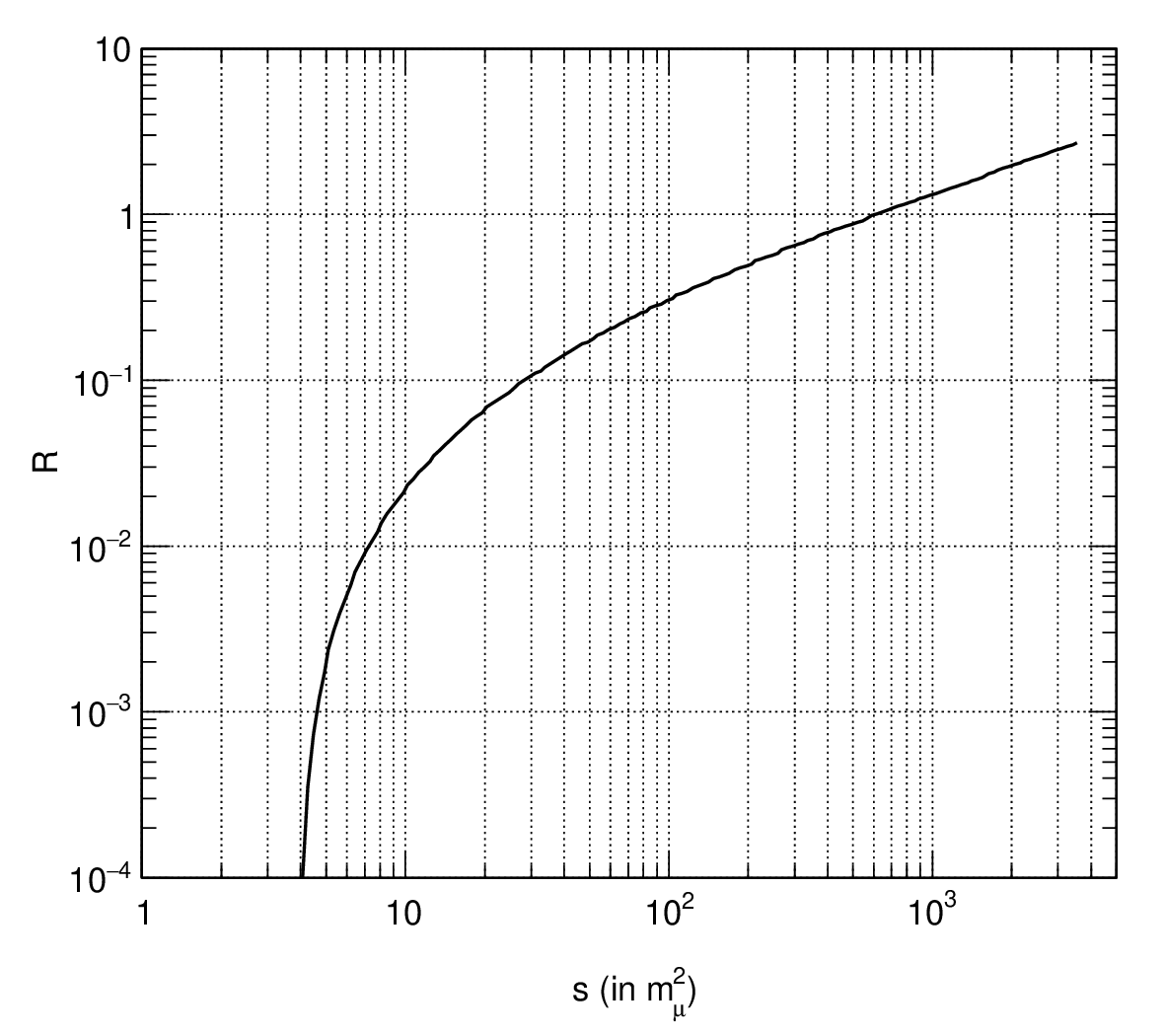}
\end{center}
\caption{The ratio $R$ as a function of $s$.}
\label{Fig-1}
\end{figure} 

The mean free path for muon pair production
\begin{eqnarray} 
l(r)=\left(\sigma_{MPP} n_{\gamma}(r)\right)^{-1}
\end{eqnarray}
where r is the radial distance from the electron acceleration region, $\sigma_{MPP}$ is the muon pair production cross section, $n_{\gamma}(r)$ is the density of photon at r. The $\sigma_{MPP}$ is around $0.1$ $\mu$b at $s \sim 20m_{\mu}^{2}$ which increases to around $0.46$ $\mu$b for $s=10^6 m_{\mu}^2$ \cite{ab4}.

For an astrophysical object of temperature T, the radiation field density is given by \cite{Rybicki2004, Link2005}
\begin{eqnarray}
 n_{\gamma}  = (a/2.8k)([1+z_{g}]T)^3 \sim 9 \times 10^{19} \left(\frac{T}{0.1 \; keV\;}\right)^{-3} cm^{-3} .
\end{eqnarray}
where $z_{g}$ is the gravitational redshift, $a$ and $k$ are the radiation and Boltzmann constant respectively. This photon density remains nearly constant over the region of interest -for example, the broad-line region of AGNs has approximately uniform temperature. We assume that the energetic electron remains within such a region until the production of a muon pair. Hence, we take $n_{\gamma}(r) = n_{\gamma}$.   
For large s ($s>> m_{\mu}^2$), the mean free path reduces to
\begin{eqnarray}
l \sim 1.7 \times 10^{12} \left(\frac{T}{0.1 \; keV\;}\right)^{-3} \left[ln \left(\frac{s}{m_e^2} \right) \right]^{-1} cm,
\end{eqnarray} 

For thermal photons with energies of $0.1$~keV and $1$~keV, the mean free paths are $1.24 \times 10^{11}$~cm and $1.24 \times 10^{10}$~cm, respectively, for $s = 20 m_{\mu}^{2}$ (corresponding to primary electron energies of 2~PeV and 200~TeV). At higher electron energies, the mean free path becomes even smaller. These values are well within the characteristic radial extents of various astrophysical sources—such as SNRs, AGNs, pulsar wind nebulae, and GRBs—as well as of specific regions within them, including the broad-line or accretion zones of AGNs. This also supports our assumption of a constant photon number density ($n_{\gamma}$) when estimating the mean free path. Finally, it should be noted that the interaction length for TPP is much smaller than that for MPP due to the significantly larger TPP cross-section.

Modeling the energies and momenta of the three final-state particles in the MPP process is highly complex and would require a detailed Monte Carlo treatment. Nevertheless, an order-of-magnitude estimate of the energies of neutrinos produced via MPP can be obtained. The inelasticity of the MPP reaction ($\eta_{\mathrm{MPP}}$) in the center-of-mass (CM) frame, in the large-$s$ limit, is approximately $\eta_{\mathrm{MPP}} \sim 3.44 (s/m_{\mu}^{2})^{-0.5}$ \cite{ab4}. For a source temperature of $0.1$~keV, the resulting muon energies are expected to be on the order of several hundred TeV for PeV electrons. Upon muon decay, its energy is, on average, distributed nearly equally among the three secondary leptons, including one electron neutrino and one muon neutrino. Consequently, the resulting neutrinos are expected to have characteristic energies in the range of tens to hundreds of TeV.

Having established the feasibility of producing high-energy neutrinos through leptonic processes, the next key question concerns the flux of such neutrinos. If the flux is too low, their astrophysical significance would be limited. The peak cross-section for $p\gamma$ interactions near the $\Delta$-resonance is approximately $500~\mu$b, whereas the MPP cross-section is about $0.1~\mu$b for $s \sim 20 m_{\mu}^{2}$. Based solely on the interaction cross-sections, the neutrino flux from leptonic processes would therefore appear three to four orders of magnitude smaller than that from $p\gamma$ interactions, assuming comparable primary electron and proton fluxes. However, the situation changes significantly when the effective interaction cross-section is considered, as discussed below.

Above the electron–positron pair-production threshold, the dominant processes governing the interaction of electrons with ambient photon fields are TPP \cite{ab12, ab12a} and IC scattering. In the TPP reaction, the fractional energy loss per interaction is small, making IC scattering the dominant cooling channel for electrons at moderate energies. At very high energies, however, the IC cross-section decreases sharply in the deep Klein–Nishina regime, approximately as $\ln(2\epsilon^{\prime})/\epsilon^{\prime}$, where $\epsilon^{\prime} \equiv \epsilon_{e}\epsilon_{\gamma}/m_{e}$ is the photon energy in the electron’s rest frame. In contrast, the cross-sections for TPP (and for MPP/PPP) remain roughly logarithmically constant. When $\sqrt{s} > 35 m_{e}$, energy losses from TPP exceed those from IC scattering \cite{ab13,ab13a}. At even higher energies, particularly for $\sqrt{s} > 20 m_{\mu}$, the MPP cross-section $\sigma_{\mathrm{MPP}}$ also surpasses $\sigma_{\mathrm{IC}}$. Consequently, MPP and PPP interactions become important energy-loss mechanisms for electrons at such extreme energies.

The ratio of the MPP to TPP cross-sections at large $s$ is approximately $2m_{e}^{2}/m_{\mu}^{2}$ \cite{ab4}, implying that out of $10^{5}$ $e+\gamma$ interactions producing triplet pairs, only about four result in MPP. However, the inelasticity of the TPP reaction, $\eta_{\mathrm{TPP}}$, is very small; in the large-$s$ limit, $\eta_{\mathrm{TPP}} \sim 3.44 (s/m_{e}^{2})^{-0.5}$ (in the CM frame). This means that one of the electrons produced in TPP retains most of the incident electron’s energy—the so-called “leading” electron—which can continue to interact with the ambient photon field until its energy falls below the threshold for the MPP. During each such interaction, there remains a finite probability for an MPP or PPP event to occur. As a result, the cumulative probability of producing muons or pions via these processes is significantly enhanced, effectively increasing the effective MPP/PPP cross-section to values comparable to $\sigma_{\mathrm{TPP}}$. When this enhancement due to the small inelasticity of TPP is taken into account, high-energy leptonic neutrino production becomes comparable to that from $p\gamma$ interactions, as demonstrated by the simulation results described below.

To quantitatively evaluate the flux and spectral distribution of neutrinos produced through leptonic channels, we performed detailed Monte Carlo simulations, as outlined below. We consider three different radiation field temperatures: $0.01$, $0.1$, and $1$~keV. Electrons are allowed to undergo TPP, IC scattering, MPP, and PPP through interactions with the ambient radiation field. The respective probabilities are determined from the interaction cross-sections given in Eqs.~(\ref{eq1}); for IC, we adopt Eq.~(8) of \cite{Lyutikov13}. The same treatment applies to secondary electrons as long as their energies remain above the MPP threshold.

The total inelastic cross-section for the $e+\gamma$ interaction is
\[
\sigma = \sigma_{\mathrm{IC}} + \sigma_{\mathrm{TPP}} + \sigma_{\mathrm{MPP}} + \sigma_{\mathrm{PPP}},
\]
where $\sigma_{\mathrm{IC}}$ is the inverse Compton cross-section. At each interaction, fractions $\sigma_{\mathrm{TPP}}/\sigma$, $\sigma_{\mathrm{MPP}}/\sigma$, and $\sigma_{\mathrm{PPP}}/\sigma$ are assumed to proceed through the corresponding channels. Since TPP has very low inelasticity, one of the produced electrons retains most of the initial energy and can continue to interact with the photon field until the threshold condition is no longer satisfied. Muons and pions generated in these processes subsequently decay into neutrinos.

\begin{figure}[h]
  \begin{center}
\includegraphics[scale=0.45]{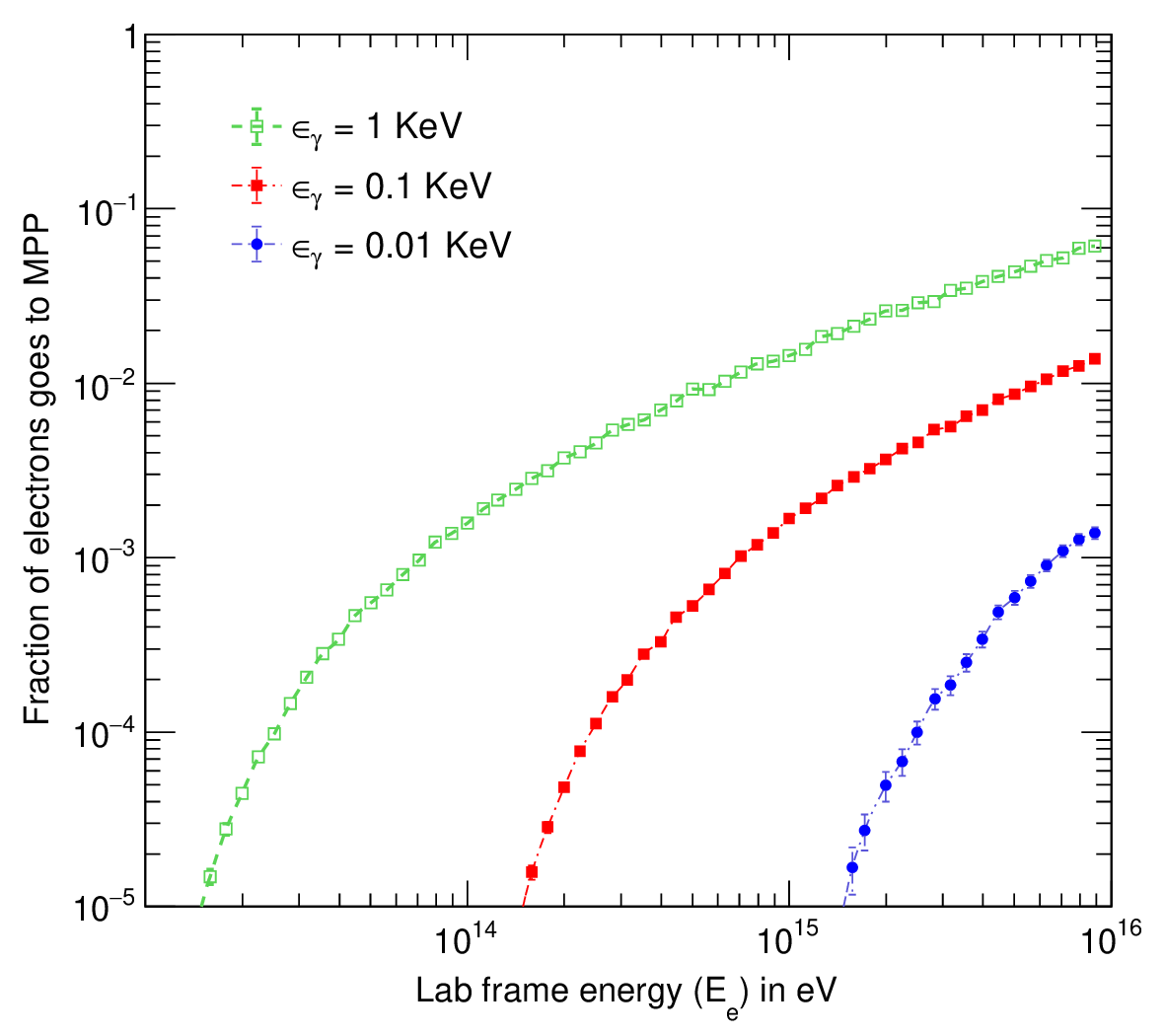}
\end{center}
\caption{The fraction of electrons that eventually produce a muon pair, rather than cooling below the threshold energy without undergoing MPP, plotted as a function of electron's lab energy.}
\label{frac_e}
\end{figure}

To estimate the fraction of primary electrons that produce neutrinos through MPP before their energies drop below threshold, we simulate $10^9$ events with electron energies between 10~TeV and 10~PeV for each seed photon energy, distributed as $N(>\epsilon_e) \propto \epsilon_e^{-1.0}$. The resulting fractions of electrons undergoing MPP interactions for photon energies of 0.01, 0.1, and 1~keV are shown in Fig.~\ref{frac_e}.

The neutrino energy distribution per event, $\frac{dN_{\nu}}{dx}$, is evaluated following Kelner et al. (2006) \cite{Kelner06}, where $x=\epsilon_{\nu}/\epsilon_e$. Adapting the formalism of H$\ddot{u}$mmer et al. (2010) \cite{Hmmer2010}, the response function for $e\gamma$ interactions under head-on approximation is
\[
R_{\nu}(\epsilon_{\nu},\epsilon_e) = \sigma \frac{dN_{\nu}}{dx},
\]
and the neutrino emissivity for a unit photon density of energy $\epsilon_{\gamma}$ is
\[
Q_{\nu}(\epsilon_{\nu}) = \int R_{\nu}(\epsilon_{\nu},\epsilon_e) \frac{dn_e}{d\epsilon_e} \frac{d\epsilon_e}{\epsilon_e},
\label{qq}
\]
where $\frac{dn_e}{d\epsilon_e} \propto \epsilon_e^{-\alpha}$ is the electron spectrum with $\alpha=2$, consistent with diffusive shock acceleration. The corresponding $p\gamma$ neutrino emissivity is obtained analogously using the analytical expression of \cite{Kelner08}, assuming the same primary particle spectrum between 10~TeV and 10~PeV.

The fractions of primary electrons undergoing MPP interactions with photon fields of these characteristic energies are shown in Fig.~\ref{frac_e}. 

\begin{figure}[h]
\begin{center}
\includegraphics[scale=0.45]{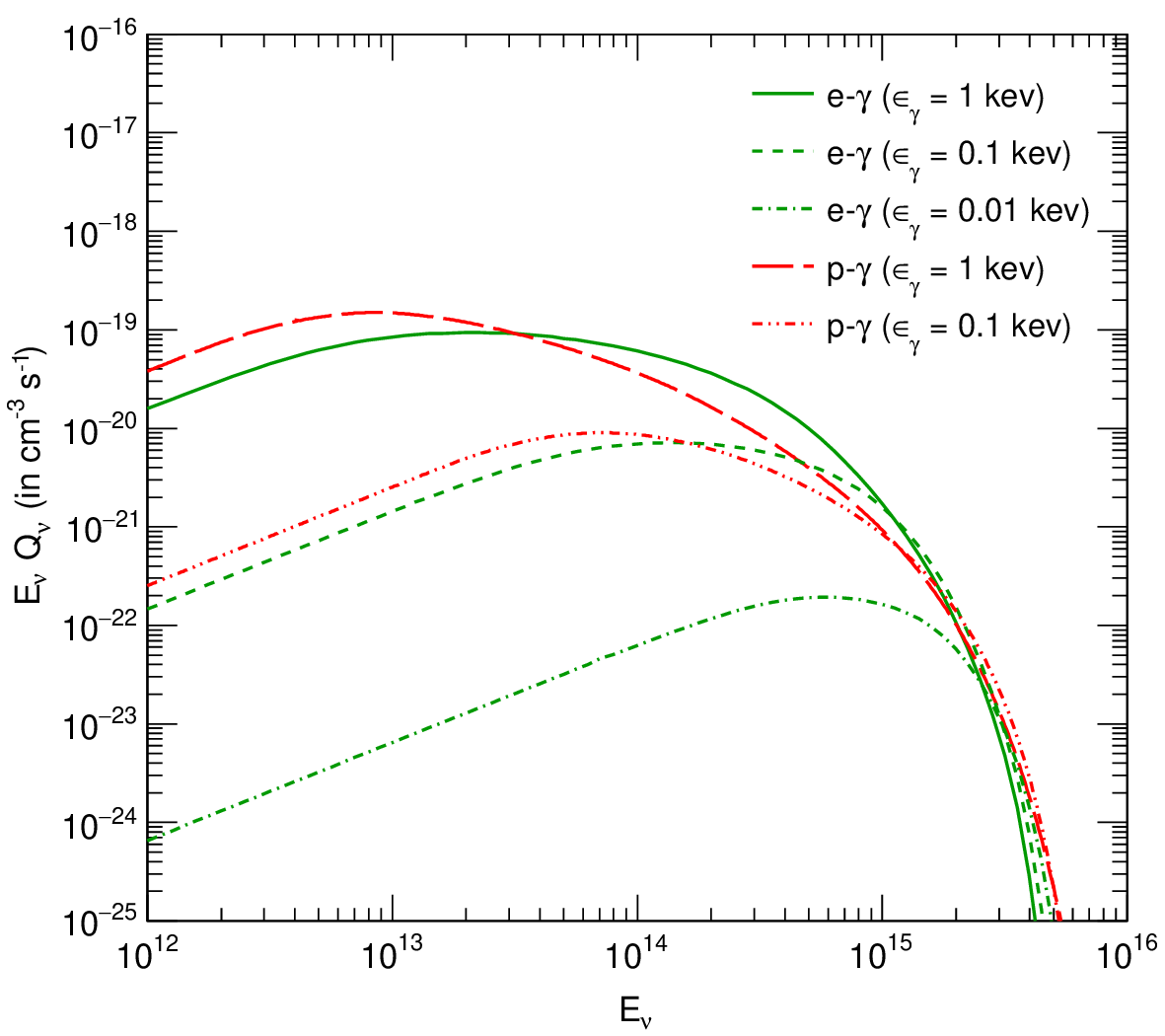}
\end{center}
\caption{Comparison of the emissivity (the rate at which neutrinos are produced per unit volume, per unit time, and per unit energy) of neutrinos produced via leptonic and hadronic processes for three different radiation field temperatures: $0.01$, $0.1$ and $1$ keV, of the radiation field of the source. The energy density of electrons/protons in the source is assumed to be $1$ erg/$cm^3$}
\label{Fig:2}
\end{figure}

The neutrino emissivities corresponding to seed photon energies of $0.01$, $0.1$, and $1$ keV are shown in Fig.~\ref{Fig:2}, assuming an energy density of relativistic electrons or protons in the source of $1$ erg cm$^{-3}$. This corresponds to an electron (or proton) luminosity of $3.7\times10^{43}$ erg s$^{-1}$ for an emission region of radius $10^{16}$ cm. Since the cross-section for $p\gamma$ interactions declines sharply beyond the $\Delta$-resonance peak, the hadronic neutrino spectra plotted in Fig.~\ref{Fig:2} represent an upper limit to the expected emissivity. We have not included TeV neutrinos arising from $pp$ collisions, as the ambient matter density in the source environment is currently unconstrained. A very recent study by Esmaeili et al. (2024) \cite{Esmaeili2024} reported that Athar et al. \cite{ab4} overestimated the MPP inelasticity at high energies. As our simulations focus on the near-threshold regime, where the updated values are $\sim30\%$ lower, adopting their correction would slightly increase the neutrino energies but would not materially affect the resulting electron-induced neutrino spectrum.

The threshold energy for the $\Delta$-resonance in $p\gamma$ interactions is roughly an order of magnitude higher than that for the MPP reaction. In the present scenario, the effective cross-section of the MPP process is approximately one to two orders of magnitude smaller than that of the TPP process, and is therefore comparable to that of $p\gamma$ interactions. The inelasticity of the $p\gamma$ reaction is higher and remains nearly constant across the relevant energy range \cite{Hmmer2010, Mucke99}. Consequently, the emissivities of hadronic and leptonic neutrinos are found to be comparable, differing by only a factor of two to three when the input fluxes of protons and electrons are assumed to be equal. For a radiation field of 0.1~keV, the hadronic neutrino emissivity is roughly twice that of leptonic neutrinos at $E_\nu \sim 1$~TeV. The two become comparable near 100~TeV, with the hadronic component decreasing by about a factor of two around 800~TeV, before converging again at 3–5~PeV. For a radiation field of 1~keV, the leptonic and hadronic neutrino emissivities remain close—within a factor of two to three—over the entire energy range, except above 3~PeV, where the leptonic component drops sharply due to the $s$-dependence of the inelasticity in the MPP reaction.  

\section*{Discussion}
It is worthwhile to mention that earlier non-hadronic neutrino production primarily via $\eta$-resonance formation in various astrophysical sources was discussed \cite{ab15}. Recently, another study \cite{Hooper23} proposed that very high-energy synchrotron photons, when scattered with X-rays and exceeding the threshold for muon–antimuon pair production, may give rise to non-hadronic neutrinos. Such neutrinos would result from a two-step process: first, energetic electrons produce high-energy synchrotron photons; in the second stage, these photons scatter with X-rays. The threshold condition for muon-pair production requires that synchrotron photons reach TeV energies when interacting with thermal photons in the keV range. To accelerate electrons to PeV energies via diffusive shock acceleration, the magnetic field must be on the order of milli-Gauss due to synchrotron energy losses. In such a magnetic field, the energy corresponding to the critical synchrotron frequency lies in the sub-GeV range. Above this critical frequency, the number of synchrotron photons decreases exponentially. As a result, the total number of synchrotron photons above TeV energies emitted by a PeV electron traveling a distance of  $10^{12}$ cm in a sub-Gauss magnetic field is effectively zero. To address this limitation, Hopper and Plant \cite{Hooper23} proposed that if the energetic electron moves into a region with a much stronger magnetic field after acceleration—such that the critical synchrotron frequency shifts to higher energies—the resulting synchrotron emission could produce an effective neutrino flux. However, this scenario requires specific (uncommon) physical conditions.

To address the question of identifying the origin of IceCube detected neutrinos, we propose that a combined study of high-energy neutrinos and the accompanying high-energy gamma rays from an astrophysical source would be valuable. In the leptonic scenario, the energetic gamma-ray flux from such sources is expected to consist of upscattered photons produced via the inverse Compton process. Since electrons undergo several TPP reactions before their energy drops below the threshold for the reaction, gamma rays may also be generated through pair annihilation. However, the flux of these gamma rays will be much smaller compared to the inverse Compton scattered photons, as the number density of electron-positron pairs is many orders of magnitude lower than the photon density in the source. The observations of TeV gamma rays from sources such as AGNs, GRBs, and SNRs are often explained as resulting from inverse Compton scattering of the source's radiation field by energetic electrons. As previously noted, the inverse Compton cross-section decreases sharply with increasing energy, leading to a rapid decline in TeV gamma-ray flux at higher energies. Consequently, if energetic electrons are responsible for both the observed gamma rays and neutrinos from a source (when the proposed neutrino sources are not optically thick to gamma rays), the ratio of high-energy neutrino flux to gamma-ray flux would increase with energy. In contrast, for gamma rays and neutrinos originating from hadronic processes, the gamma ray and neutrino fluxes would be approximately of the same order \cite{Kelner06} and remain almost constant over a broad energy range.

In estimating the diffuse neutrino flux, as well as neutrino fluxes from celestial objects, contributions from pp and p$\gamma$ interactions are typically considered. The interstellar radiation field and interstellar gas serve as targets for the production of diffuse gamma rays by energetic hadronic cosmic rays. In light of the present findings, the potential role of energetic electrons in generating TeV neutrinos through interactions with the Galactic ultraviolet and soft X-ray background radiation also warrants careful evaluation.
The electromagnetic spectrum of a typical AGN usually exhibits two peaks, one at around a few eV (synchrotron peak) and the other at around 100 MeV. The higher energy peak is generally described in terms of IC scattering of primary relativistic electrons with synchrotron photons comoving with the AGN jet. 
For the interaction of PeV electrons with the IC scattered (MeV) photons the square of the centre-of mass energy will be $s \sim 10^6 \; m_{\mu}^2$. At such high energies, the MPP process is expected to play a significant role. 
One of our main objectives is to explore whether at least some of the PeV neutrinos detected by the IceCube Observatory could have a leptonic origin. Photohadronic ($p\gamma$) production of PeV neutrinos has generally been proposed in the literature to explain the IceCube observations \cite{Ansoldi18,Gao19}. However, as we have shown, electrons accelerated to PeV energies can also produce a neutrino flux comparable to that generated by protons with the same primary flux interacting with the ambient photon field. We are currently investigating whether the proposed leptonic models can consistently explain the neutrino observations from any of the IceCube point sources, along with the corresponding electromagnetic spectra.

\section*{Conclusion}

Our results demonstrate that high-energy neutrinos can arise not only from hadronic interactions but also from purely electromagnetic processes involving extremely energetic electrons. This finding challenges the long-held assumption that the detection of TeV–PeV neutrinos necessarily implies hadronic acceleration at the source. The comparable flux levels achievable through leptonic channels suggest that energetic electrons may contribute significantly to the neutrino population observed by IceCube. This realization underscores the importance of treating leptonic and hadronic scenarios on an equal footing when interpreting high-energy neutrino observations.

A decisive distinction between these two production channels requires coordinated multi-messenger observations across the gamma-ray and neutrino domains. Future measurements extending over a broad energy range—from GeV to PeV energies—by facilities such as the Cherenkov Telescope Array (CTA), IceCube-Gen2, and KM3NeT, will be crucial in identifying the true origins of high-energy neutrinos. By establishing whether the underlying mechanisms are hadronic or leptonic, such studies will not only refine our understanding of cosmic-ray acceleration but also deepen insight into the most extreme environments in the Universe.



\end{document}